\documentclass{osa-article}

\journal{osajournal}

\articletype{Research Article}

\usepackage{subcaption}
\usepackage{physics}
\usepackage{hyperref}
\usepackage{mathtools}
\newcommand{\tem}[1]{HG$_{#1}$}
\newcommand{\temlg}[1]{LG$_{#1}$}
\definecolor{emerald}{rgb}{0.314, 0.784, 0.471}

\usepackage{lineno}

\begin{document}

\title{Differential wavefront sensing and control using radio-frequency optical demodulation}

\author{Daniel Brown\authormark{*}, Huy Tuong Cao, Alexei Ciobanu, Peter Veitch, David Ottaway}

\address{School of Physical Sciences, University of Adelaide, Adelaide, 5005, Australia \\
Ozgrav-Adelaide, Australian Research Council Centre of Excellence for Gravitational Wave Discovery, Australia
}

\email{\authormark{*}daniel.d.brown@adelaide.edu.au} 

\pagestyle{plain}
\begin{abstract}
Differential  wavefront  sensing  is  an  essential technique  for  optimising the performance of many precision interferometric experiments. Perhaps the most extensive application of this is for alignment sensing using radio-frequency beats measured with quadrant photodiodes. Here we present a new technique that uses optical demodulation to measure such optical beats at significantly higher resolutions using commercial laboratory equipment. We experimentally demonstrate that the images captured can be digitally processed to generate wavefront error signals and use these in a closed loop control system for correct wavefront errors for alignment and mode-matching a beam into an optical cavity to 99.9\%. This  experiment   paves  the  way  for  the  correction of even higher order errors when paired with higher order wavefront actuators. Such a sensing scheme could find use in optimizing complex interferometers consisting of coupled cavities, such as those found in gravitational wave detectors, or simply just for sensing higher order wavefront errors in heterodyne interferometric table-top experiments.
\end{abstract}

\section{Introduction}

Wavefront aberrations in optical fields can limit the performance of state of the art experiments in gravitational wave interferometry (GW)~\cite{Mavalvala:98, Brooks:16, Sasso_2018}, quantum optics~\cite{PhysRevLett.121.263602}, vortex beam generation~\cite{Ohland}, and atom interferometers~\cite{miga:18, PhysRevA.93.043610, PhysRevApplied.7.034016, Schkolnik}. Aberrations can reduce the sensitivity of an experiment through lost optical power or increasing the coupling of noise sources to signal readouts. In particular, gravitational wave interferometers require continuous sensing and control of suspended optics to reduce alignment wavefront errors. This optimizes the stored optical power within their cavities and reduces alignment noise that degrades their sensitivity. With the ever increasing push for improved sensitivity to astrophysical events, the next-generation of detectors~\cite{Adhikari_2020, NEMO, ET, miga:18} will all require improvements to both the sensing and control of wavefronts aberrations beyond alignment to realise their full potential.

Radio-frequency (RF) differential wavefront sensing~\cite{Morrison:94:theory, Meshksar21} (DWS) is an  optical heterodyne technique for measuring the relative wavefront differences between different frequency components in a beam---typically this is between a carrier and its RF modulated sidebands. These optical beats are most often used for length sensing, like the ubiquitous Pound-Drever-Hall~\cite{PDH:84} (PDH) scheme. Demodulating these same RF optical beat over a segmented quadrant~\cite{Kawabe:94,Morrison:94:exp,PhysRevD.100.102001,PhysRevApplied.14.054013} or more specialised \textit{bullseye}\cite{Mueller:00} photodiode we can extract low-order wavefront deformations for alignment or beam size differences.

Experimental sensing and control of first order alignment and translation errors is relatively well established compared to sensing of higher order effects---such as the shape of the optical field~\cite{Mueller:00, Hechenblaikner:10}. Interest in the latter has recently increased~\cite{PhysRevD.100.102001, Ciobanu:20} for reducing losses that cause decoherence in the squeezed light states used for improving the sensitivity of GW detectors~\cite{PhysRevD.102.062003}. Recent work has realised 6~dB of improvement in the sensitivity of GEO600~\cite{PhysRevLett.126.041102}; optical losses of 6\% still remain due to wavefront errors, of which around 1\% is attributable to astigmatism whilst the rest is unknown. Sensing of astigmatism and other higher order aberrations will be required to diagnose such issues and reduce their effects, this will be especially the case in third generation detectors that will require up to 10~dB of squeezing. 

Simply increasing the number of segments up from quadrant photodiodes can be challenging due to crosstalk between the elements at radio frequencies~\cite{10.1117/12.2015609}. CCD cameras have been used for imaging 10kHz heterodyne beats~\cite{Cervantes:07} and a prototype CMOS camera with on board per-pixel RF demodulation has been demonstrated at the prototype level by Patel~\textit{.et.al}\cite{Patel:11}---although the latter devices do not appear to be commercially available.
Time-of-flight cameras~\cite{Luan:01} may also offer a potential solution as these devices can perform on board demodulate at RF frequencies; although finding one suitable for use in interferometric experiments challenging due to them being highly specialised for depth sensing applications.
Mechanical scanning devices~\cite{Goda:04, Agatsuma:19} and digital interferometry techniques~\cite{Tarquin_Ralph_2020} have also been developed to image at higher resolutions than simple segmented photodiodes. Another possible technique which we develop further in this work is using \textit{optical demodulation}.
By moving to optical demodulation we can remove mechanical features that limit framerate, resolutions and any potential vibration couplings. Recently an all-optical device operating at kilohertz frequencies was  demonstrated by Panigrahi~\textit{.et.al}~\cite{Panigrahi:20}. As optical demodulation can be achieved using relatively standard commercial equipment, it offers an alternative to specialist and complex camera based sensors with integrated RF electronics which may not be suitable for specific wavelengths, especially at anything longer than 1um. 

The optical lock-in camera~\cite{Cao:20p} (OLC) technique demonstrated here performs optical demodulation at megahertz frequencies. High resolution images are digitally processed to mimic physically segmented photodiodes for simple data extraction. They can provide diagnostics of control signals generated from physical segmented photodiodes by quantify deformations and higher order mode content in the signal. Being digital, the segmentation used can be arbitrarily complex and be altered during operation offering far more adaptability over their physical counterparts. Alternatively, with higher resolutions available advanced analysis pipelines---such as a neural networks~\cite{10.1117/1.OE.59.7.074107}---could enable more efficient information extraction.

In this paper we demonstrate a technique using OLCs that can match a beam to a cavity in alignment and shape to 99.9\% using the same optical frequencies used for PDH locking. In section 2, we briefly review the production of the heterodyne beat when a phase-modulated beam is incident onto a cavity and how to extract information from it. In Section 3 we describe our proof-of-principle experiment that demonstrates the sensing of the misalignment, mode-mismatch and using this to form a closed-loop control system. The results of the experiment are then presented in Section 4.

\section{Using optical lock-in cameras for differential wavefront sensing}

\begin{figure}
    \centering
    \includegraphics[width=0.6\textwidth]{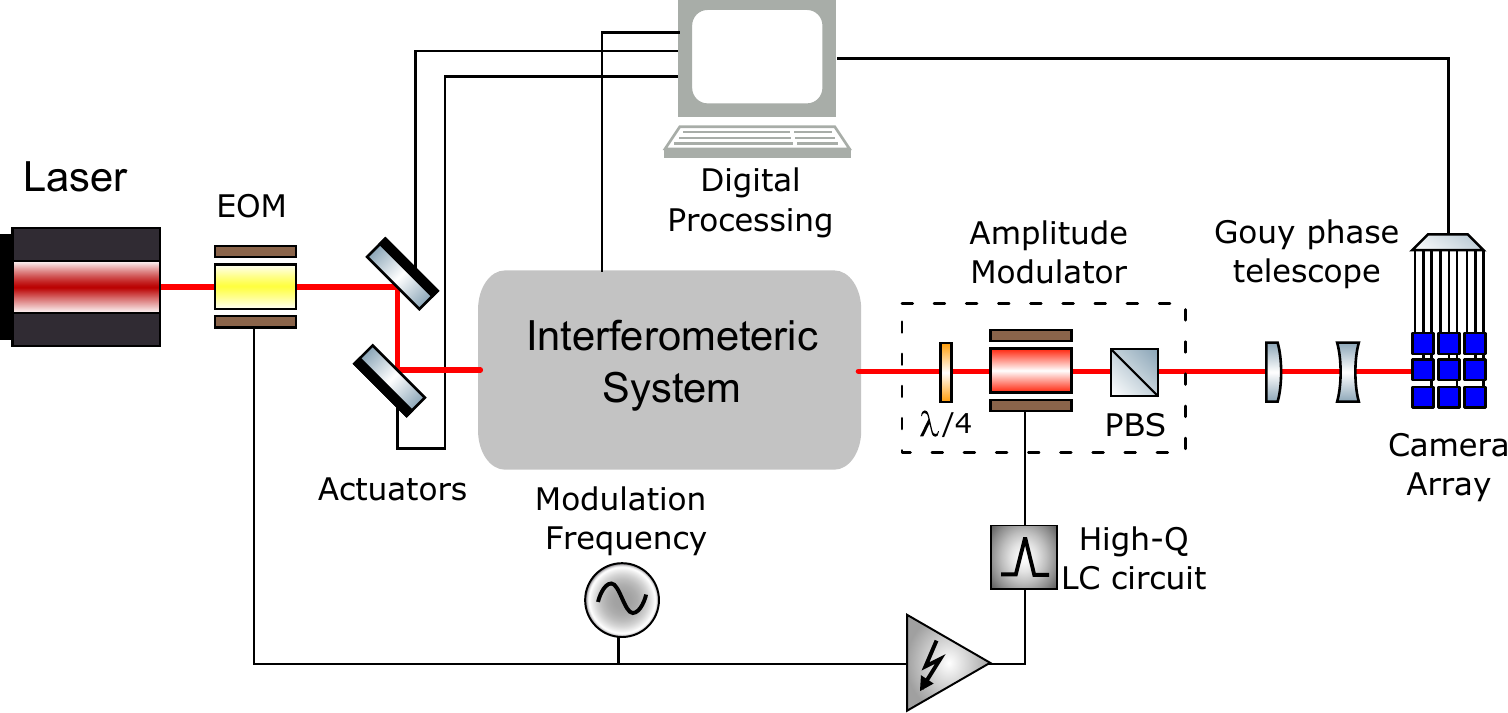}
    \caption{Schematic of the optical lock-in camera scheme. A phase-modulated beam interacts with some interferometric system, be it a simple cavity or more complex system. At an output port the beam is demodulated optically using an amplitude modulator ($\lambda/4$ waveplate, a pockels cell, and a polarising beamsplitter). The  demodulated field is then imaged and processed, the resulting information then used for various actuating inputs.}
    \label{fig:phase_camera_simplfied}
\end{figure}

Our optical lock-in camera consists of an amplitude modulator and a high-resolution camera, as shown in Fig.~\ref{fig:phase_camera_simplfied}. The amplitude modulator acts as a fast switch for demodulating an optical beat over the mega-pixel array, the array itself acts as a low-pass filter. This allows us to measure the spatially varying amplitude and phase of the optical beat. 
In this section we examine how the optical lock-in camera can be used to image wavefront errors after they have interacted with an interferometric system. Such a system could be something as simple as reflection from a Fabry-Perot cavity or more complex systems such as those found in gravitational wave interferometers; in this work we consider the behaviour for the former simple case.

Consider a linearly polarised beam incident on an optical cavity, $E_i(x, y)$. This field can be represented in many bases such as the Hermite- or Laguerre-Gaussian modes---the exact choice of which does not matter here. The incident beam is phase-modulated at a frequency $\omega$ relative to the optical frequency $\omega_0 = c/\lambda$ by an electro-optic-modulator (EOM). The field reflected from the cavity input coupled mirror, $E_r$, is then used for Pound-Drever-Hall (PDH) locking and the sensing of differential wavefront errors in the incident field relative to the cavity. We assume, for simplicity, that none of the incident fields at the sideband frequencies $\omega_0\pm\omega$ resonate in the cavity and are promptly reflected. When the cavity length is on resonance for the carrier field the reflected field will be of the form
\begin{eqnarray}
    E_r &=& \left([E_d(x,y) + \alpha E_c(x,y)] + i \frac{m_1}{2} E_i(x,y)\left[e^{i\omega t} + e^{-i\omega t}\right]\right)e^{-i\omega_0 t}.
    \label{eq:Er}
\end{eqnarray}
Here $m_1$ is the EOM modulation index and assumed to be small, $\alpha$ represents the portion of the resonating cavity field that leaks out, $E_c$, and $E_d$ is part of the incident light at the carrier frequency that has been rejected from the cavity due to wavefront distortions. On resonance $\alpha$ is real-valued, its magnitude and sign depends on whether the cavity is over-, under-, or critically-coupled. 

The reflected field is then optically demodulated at $\omega$ by the amplitude modulator. We simplify the polarisation effects here for brevity\footnote{A more detailed treatment with polarisation effects can be found in~\cite{cao_thesis}} and reduce the amplitude modulator to a simple modulation function $M(\phi) = \frac{1}{2}[1 + m_2\cos(\omega t + \phi)]$, where $m_2$ is the modulation index and $\phi$ the demodulation phase. An image of the demodulated beam is capture by a sCMOS camera over an exposure time $\tau$. This effectively low-passes any intensity fluctuations at a frequency $\propto 1/\tau$. Here we approximate a measured pixel value over a sensor area $A$ by considering the time-averaged signal
\begin{eqnarray}
    V_\phi &=& \left\langle\iint_{A}|M(\phi) E_r|^2 dA\right\rangle, \\
           &\approx& \frac{1}{2} \iint_{A} 
           \underbrace{\left(1-m_1^2/4\right)\left|\alpha E_c(x,y) + E_d(x,y)\right|^2 + \frac{m_1^2}{4} |E_i|^2}_{\text{Field intensity terms}} \nonumber \\
           & & \hspace{1cm}+ 
           \underbrace{m_1 m_2\cos(\phi) \mathrm{Re}\left\{i E_i(x,y) E^\ast_d(x,y) + i \alpha E_i(x,y) E^\ast_c(x,y)\right\}}_{\text{Optical beat terms}} \,\mathrm{d}A.\label{eq:V_ac_dc}
\end{eqnarray}
There are two sets of terms, those that vary with demodulation phase and those due to the intensities of each part of the optical field.
A more complete picture that considers the exposure time is discussed in the accompanying supplemental material.

By taking images at four different demodulation phases $\phi=\{0, \pi/2, \pi, 3\pi/2\}$ with Eq.\ref{eq:V_ac_dc} we can extract the \textit{in-phase}, $\mathbf{I}$, signal which yields a term proportional to the differential wavefront, $E_d(x,y)$,
\begin{equation}
    \mathbf{I} \equiv V_0 - V_\pi = m_1 m_2 \iint_{A}
    \Re\{
        \underbrace{i E_i(x,y) E^\ast_d(x,y)}_{
            \text{Signal}
        } +
        \underbrace{i \alpha E_i(x,y) E^\ast_c(x,y)}_{
            \text{Cavity field beat}
        }
        \} dA.
    \label{eq:I}
\end{equation}
The cavity field beat term is the PDH signal, this will only be suppressed when the cavity is locked to the incident field with no detuning present. In a more complex optical systems, such as gravitational wave interferometers, the \textit{quadrature} signal can also be found using $\mathbf{Q} \equiv V_{\pi/2} - V_{3\pi/2}$. In simple cases with two balanced sidebands you can always choose an offset for the demodulation phases such that $\mathbf{Q} = 0$.

\begin{figure}
  \centering
  \includegraphics[width=0.9\linewidth]{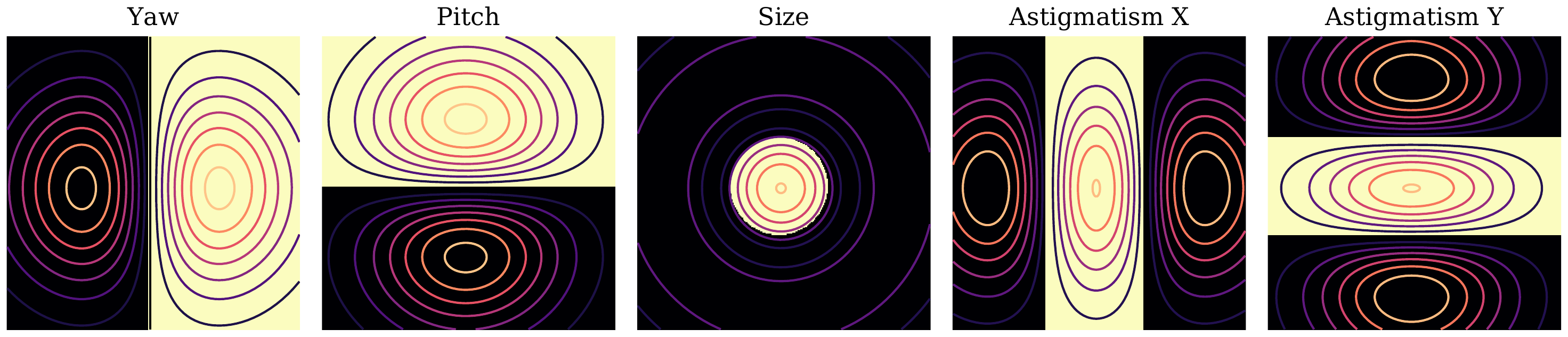}
  \caption{Shown are examples of five masks used for extracting information on the alignment and mode-matching of beams. These are overlaid with the Hermite-Gaussian modes they are targeting. Black and yellow represent the negative and positive segments when summing the image. From left to right, \tem{10}, \tem{01}, \temlg{01}, \tem{20}, and \tem{02}}
  \label{fig:output_masks}
\end{figure}

It is important to note that value $V_\phi$ is not a continuous variable, it is digitised by the camera before we perform the subtraction operations. Depending on the camera this will be represented by an $N_\mathrm{b}$-bit variable. This digital number (DN) is related to the number of photo-electrons $N_e$ by the conversion factor $K=L(N_e)\frac{2^{N_\mathrm{b}}}{W_\mathrm{d}}$, where $W_\mathrm{d}$ is the well depth of the pixel, and $L(N_e)$ is the linearity of the sensor. The $N_\mathrm{b}$-bit representation limits the smallest measurable change relative to the large DC intensities, the ratio of the field intensities and the optical beat terms in Eq.\ref{eq:V_ac_dc}, therefore cameras with high-dynamic range are preferable to maximise performance. 

\subsection{Higher order mode wavefront sensing}

Here we present an example case of sensing alignment information using the OLC. Let us consider a beam which is perfectly mode-matched but is misaligned and offset relative to the axis of an optical cavity. In the Hermite-Gaussian basis a small misalignment $\gamma$ or offset $\delta$ of a perfect \tem{00} beam will have some additional \tem{10} or \tem{01} mode. The complex valued amplitude of these higher order modes will be $\beta \propto \delta/w_0 + i\gamma/\Theta$~\cite{Anderson:84}, where $\Theta$ and $w_0$ are the divergence and waist size of the cavity resonating cavity field. Similar relationships are applicable to second-order modes and mode-matching degrees of freedom.

The optical field incident on the cavity is $E_i \propto E_d + E_c = \mathrm{HG}_{00}(x, y) + \beta\mathrm{HG}_{10}(x, y)$ ignoring some overall power scaling.
 As the \tem{00} mode is the eigenmode of the cavity, $E_c = \mathrm{HG}_{00}(x, y)$, the difference will be
$E_d = \beta \mathrm{HG}_{10}(x, y)$. From Eq.\ref{eq:I}, an anti-symmetric image will be generated by the OLC when the cavity is on resonance, $\mathbf{I} \propto \Re\{\beta\mathrm{HG}_{10}(x, y)\mathrm{HG}_{00}^\ast(x, y)\}$~\cite{Anderson:84}.
The simplest way to extract a signal relative to $\beta$ is to perform and operation similar to how a quadrant-photodiode measures the difference in power between two of its segments.
Therefore we can apply an anti-symmetric \textit{mask} to $\mathbf{I}$ and sum all the pixels to generate a single value.
In this case the mask weights all pixels to the left of centre by -1 and those to the right by +1. An example of such masks are shown in Fig.\ref{fig:output_masks} for some of the lower order spatial modes---this does not just apply to the first order modes so one in-phase image can be masked in multiple ways to extract information on different mode contents. Computationally this is a fast enough operation that it can be used to generate feedback signals.

\subsection{Gouy phase telescopes}

Given that $\beta$ is complex and we only image the real part of the optical beat we are not able to simultaneously extract the two spatial quadratures $\gamma$ and $\delta$ (alignment and translation) that a \tem{10} or \tem{01} mode can represent. Similar to how two measurements at different demodulation phases are required to extract both the in-phase and quadrature components of some oscillating signal, we must also perform two measurements at different Gouy phases; which can be thought of as a spatial demodulation phase.

When sensing specific optical modes it is critical to consider the amount of Gouy phase, $\Psi$, they have accumulated before they observed at a detector~\cite{Morrison:94:theory}. The amount of Gouy phase a non-astigmatic Hermite-Gaussian \tem{nm} mode accumulates as it propagates is $\propto (1+n+m)$. Considering our misaligned example previously, our image will evolve depending on  the amount of Gouy phase the \tem{10} accumulates whilst propagating
\begin{equation}
\mathbf{I}(\Psi) \propto \Re\{\beta\,\mathrm{HG}_{10}(x, y)\mathrm{HG}_{00}^\ast(x, y)e^{i\Psi}\},
\end{equation}
where here we have explicitly highlighted the $\Psi$ term in the HG functions.
Therefore if we take two in-phase measurements at Gouy phases separated by $\pi/2$ we can optimally extract both $\gamma$ and $\delta$---this is equivalent to measuring the beam in both the near- and far-field.

\textit{Gouy phase telescopes}~\cite{Morrison:94:theory} is name given to the lensing optics that are specifically designed to accumulate a particular amount of Gouy phase when a beam propagates through it. Two or more Gouy phase telescopes are typically required. For the first order modes the amount of accumulated Gouy phase must differ by $\pi/2$ between them. For higher orders, this would need to be less, i.e. \tem{20} would require $\pi/4$. These values however are for optimal sensing, if we want to sense both \tem{10} and \tem{20} we can simply design the telescopes for a difference of $\pi/3$. 
In Fig.\ref{fig:phase_camera_simplfied}
we place the telescope between the amplitude modulator and the sensor, however it could be beforehand if required.

\section{Description of the experiment}

\begin{figure}
    \centering
    \includegraphics[width=0.5\textwidth]{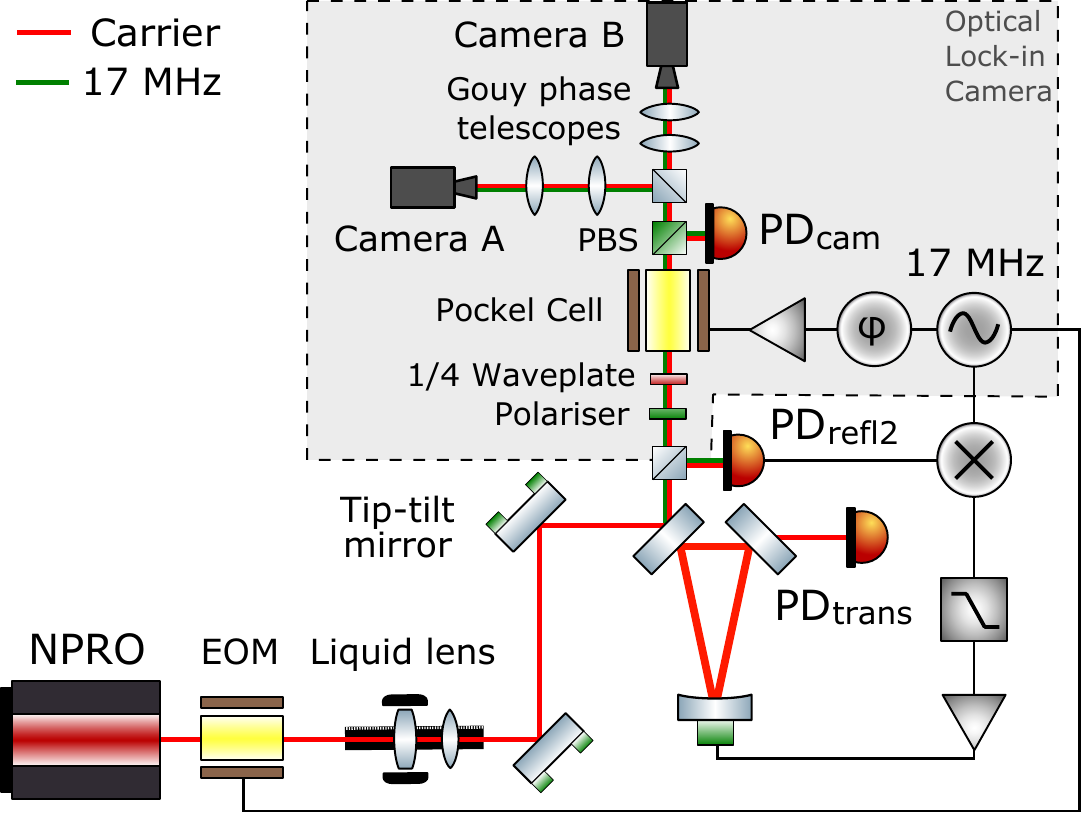}
    \caption{Experimental setup used for this work. Two tip-tilt mirrors and liquid lens actuators are used for optimising the beam coupling into the cavity shown. The optical lock-in camera is placed on reflection of the cavity for measuring relevant error signals.}
    \label{fig:experiment}
\end{figure}

The aim of our experiment was to demonstrate that the OLC images can be used to for a closed loop control of both the alignment and mode-matching of a laser beam into an optical cavity. Once these low order wavefront distortions were corrected we were then be able to see any higher spatial errors remaining. A schematic of the setup used is depicted in Fig.~\ref{fig:experiment}.

The experiment used an Innolight Memphisto NPRO laser, $\lambda=1064$nm, which was phase-modulated at 17~MHz. The optical cavity used was a triangular cavity with a finesse of $\approx 400$. The 17~MHz optical beat was measured on reflection of the cavity to generate a PDH for length locking. The digital control toolbox PyRPL~\cite{neuhaus_pyrpl_2017} on Red Pitaya StemLab 125-14 was used to digitally demodulated and filter the PDH signal. This then controlled the length of the cavity with a piezoelectric transducer on one of the cavity mirrors.

A fraction of reflected beam was also sent to the OLCs. A high extinction ratio polariser and a half-waveplate were used for intensity control and ensuring a linearly polarised field enters the OLC. The amplitude modulator was formed using a quarter-waveplate to circularly polarise the light going into a Pockels cell (PC).
The Pockels cell was electrically part of a series LC-circuit (resonant tank circuit) that was tuned to resonate at 17MHz, this provided both frequency selectivity and enhance the voltage across the cell. The Pockels cell switched  the optical field between S- and P-polarisation when driven with its quarter wave voltage. The polarizing beamsplitter (PBS) then selects only one of the polarisations to complete the amplitude modulator.

A Siglent SDG6022 was used to generate the 17~MHz signal for driving the resonant tank circuit, this signal was amplified by a Mini Circuits LZY-22+ broadband RF amplifier. To set the optical demodulation phase, the phase of the 17~MHz signal was externally shifted using the auxiliary input on the SDG6022. This phase control signal was generated by another StemLab 125-14. This StemLab also generated another signal which is used to trigger the cameras each time the demodulation phase was changed.

The OLC used two sCMOS Andor Zyla 4.2 cameras, which were chosen for their linearity, high-dynamic range, and low noise properties. Given that balanced sidebands are used in this experiment a demodulation phase was chosen so that all the signal appears in the $\mathbf{I}$ images. This meant only two images for each camera had to be captured to extract all the information. Exposure times were kept as short as possible, around 10 to 50~us, which helped to reduce the effect of stray light and movements in the lab from being picked up by the camera between the two images. The cameras ran with a cropped region-of-interest (ROI) to accelerate data capture. Camera A captured 1000x1000 images with a 4x4 binning, which camera B 500x500 and 2x2 binning. This ROI was constrained by the size of the beams on the cameras, set by our Gouy phase telescope target of $\Psi=\pi/3$ for sensing both the first and second order modes; the beam on camera A was larger than that on B. 
This all resulted in capturing $\mathbf{I}$ images at 100~Hz at 250x250 resolution. Higher resolutions at such speeds were limited by the speed of our software-based digital processing.

The alignment of the incident beam into the cavity was controlled electrically using a pair of tip-tilt mirrors (Physik Instrumente S-330 Piezo Tip/Tilt). The waist size and position were adjusted with a liquid lens (Optotune EL-10-30-C NIR with a -1 to +5D focal power range) on a manual translation stage.
This resulted in five actuator degrees of freedom that need to be controlled. There are common and differential (COMM and DIFF) pitch and yaw between the two tip-tilts, and the focal length of the liquid lens. The former set the alignment and translation of the beam into the cavity and the latter a mix of waist size and position. The liquid lens translation stage was adjusted by hand when the other loops were closed.

The signal processing of the OLC images was handled by custom Python and C software. The $\mathbf{I}$ images from each camera were masked multiple times for each degree of freedom and summed to generate error signals for the control loops. Choosing the basis (spot size and position on the CCD) to generate the masks was important, as it affects the zero-point of the error signals. The cavity was first reasonably well mode-matched and aligned by hand, then steering mirrors in front of each camera were used to center the beam in the images. The masks were then generated using the position and size of the beams.
Small changes in the masks or offsets in the error signals themselves were used to fine tune the control loops to maximise the transmitted power. 
These digital signals were then send to a digital-to-analog converter for driving the tip-tilts and liquid lens. The control loops just consisted of simple integrators. Maximising the bandwidth of these loops was not the primary target of our work, the digital latency limited us to $\approx 20~Hz$, which was adequate for our needs here.

\section{Results}

\begin{figure}[t]
  \centering
\includegraphics[width=\linewidth]{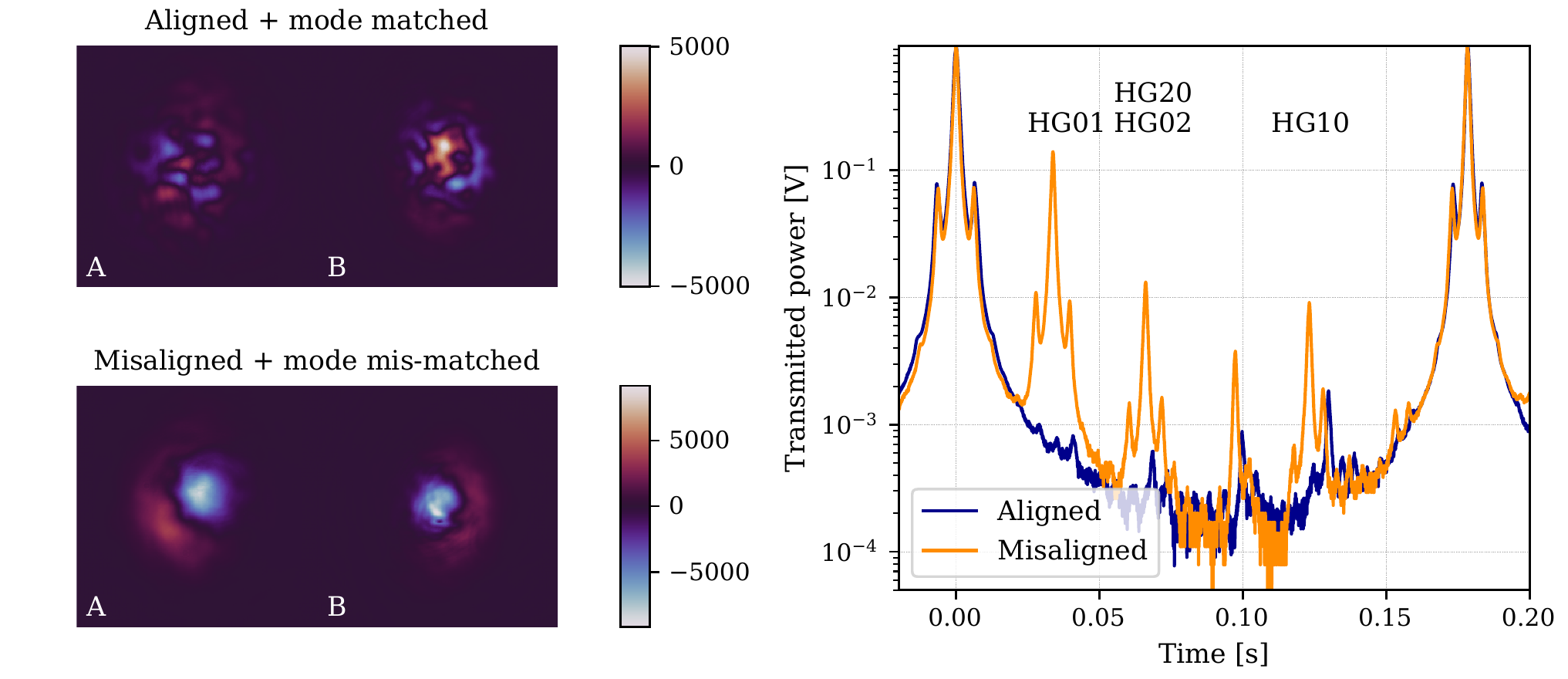} 
\caption{Images on the left show the optical lock-in camera images taken in an aligned and misaligned and mismatched state. The beam on the left is from Camera A and on the right Camera B. For comparison, the cavity mode spectrum of higher-order modes seen by the cavity for both states are shown on the right. The peak is clipped at 1V due to the maximum ADC input to provide more resolution for the scan, the peak voltage of the \tem{00} is $1.1$V.}
\label{fig:cavity_scan}
\end{figure}

A simple method used for measuring a beam's mode content relative to a cavity is by sweeping the cavity's length over a free-spectral-range whilst observing the transmitted power. Peaks in the transmission can then be matched with known higher-order mode resonances inferred from the cavity's geometry---although this is limited by modal degeneracy and sign ambiguities. We can estimate the total magnitude of effects such as misalignment and mismatch by computing the power ratio of different modes relative to \tem{00}. For example, the power lost from the \tem{00} resonance due to misalignment and offset errors can be determined by $|\beta|^2=P_{10}/P_{00}$. It is not possible to extract $\gamma$ and $\delta$ from such a measurement due to the loss of phase information from measuring just transmitted power.

The OLC images shown in Fig.\ref{fig:cavity_scan} were taken with the cavity locked to the \tem{00} mode. Keeping various actuators fixed, the cavity was quickly unlocked and a sawtooth signal applied to the cavity's piezoelectric actuator whilst measuring the transmitted optical power to perform the scan. In the misaligned case, the mode content is a mix of the first and second order modes which can be clearly seen in both the images as well as in the spectrums. The cavity was then re-locked and the control loops were closed. Once settled, the actuators were fixed, and another spectrum taken. The lower order modes are mostly suppressed now, as can be seen in the images and spectrum. Using the power ratios of the \tem{01}, \tem{10}, and \temlg{01}=\tem{20}+\tem{02} to the \tem{00}, we see the control loops have improved the alignment and mode-matching to $\approx 99.9$\%. Careful tuning of the offsets in each of the error signals was crucial to reach this optimised state.

\begin{figure}[t]
  \centering
\includegraphics[width=\linewidth]{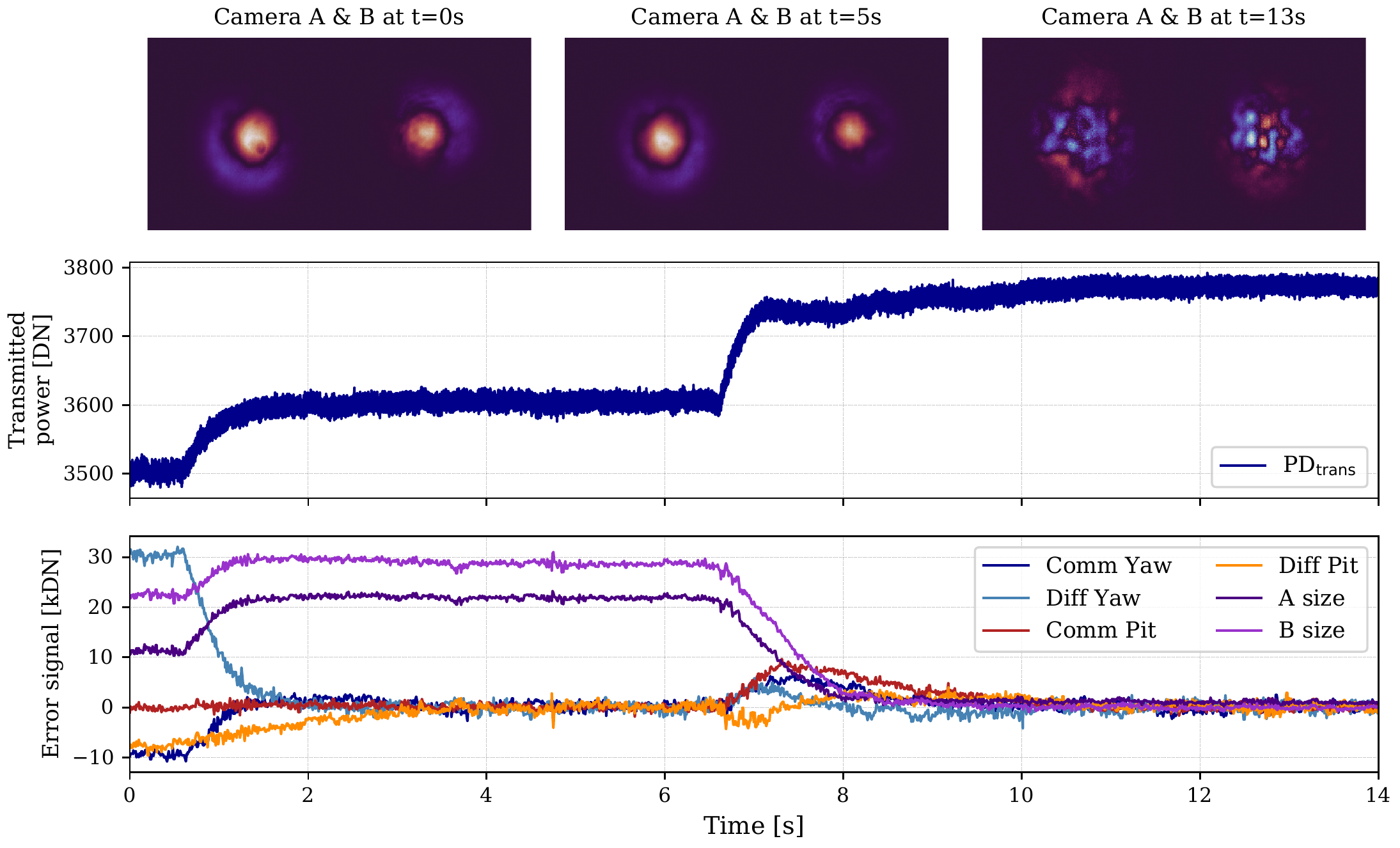} 
\caption{Automatic alignment and mode-matching optimisation. Shown  are the output images at the three different states: misaligned and mismatched, mismatched, optimised. The central plot shows the power transmitted through the cavity. The bottom plot then shows the error signals for the different degrees of freedom. Alignment loops are switched on first, around 0.3s in, and the mode-matching loops around 7s. The color scale is re-normalised for each set of images for visibility. }
\label{fig:lock_acquisition}
\end{figure}

In Fig.~\ref{fig:lock_acquisition} we show the time domain behaviour of the system automatically aligning and mode-matching the incident beam to the cavity using the OLC generated signals. At first there are clearly visible misalignment and mismatch between the two in-phase images. The alignment loops are closed first around 1s in, with differential yaw being the largest source of error. Lastly the mode-matching loops are closed at 7s. As seen in by the error signals there is some cross coupling between each of the degrees of freedom due to the liquid lens affecting the alignment of the beam. After they have settled transmitted power has been increased by $\approx 7\%$.

Our technique improves over simple segmented photodiodes by providing information on any remaining wavefront errors that exist. Shown in the images at $t=13s$ in Fig.~\ref{fig:lock_acquisition} is the left over distortions after the \tem{10}, \tem{01}, and \temlg{01} have been suppressed by our control system. The remaining field consists mainly of \tem{02} mode, or vertical astigmatism for which we have no dedicated actuator to correct with. It was possible to manually alter the alignment of the beam through liquid lens to alter this astigmatism slightly, hence the difference in the controlled images between Fig.~\ref{fig:cavity_scan} and Fig.~\ref{fig:lock_acquisition}.
There is a significant amount of high spatial frequency errors which are present due to the the imperfect optical quality of multiple components in the system. These additional higher order modes contribute to offsets in calculated error signals~\cite{sigg_wavefront}. Stationary effects can be easily compensated for with simple offsets, however any time vary effects cannot be easily distinguished from a real signal.
\begin{figure}[t]
  \centering
\includegraphics[width=0.6\linewidth]{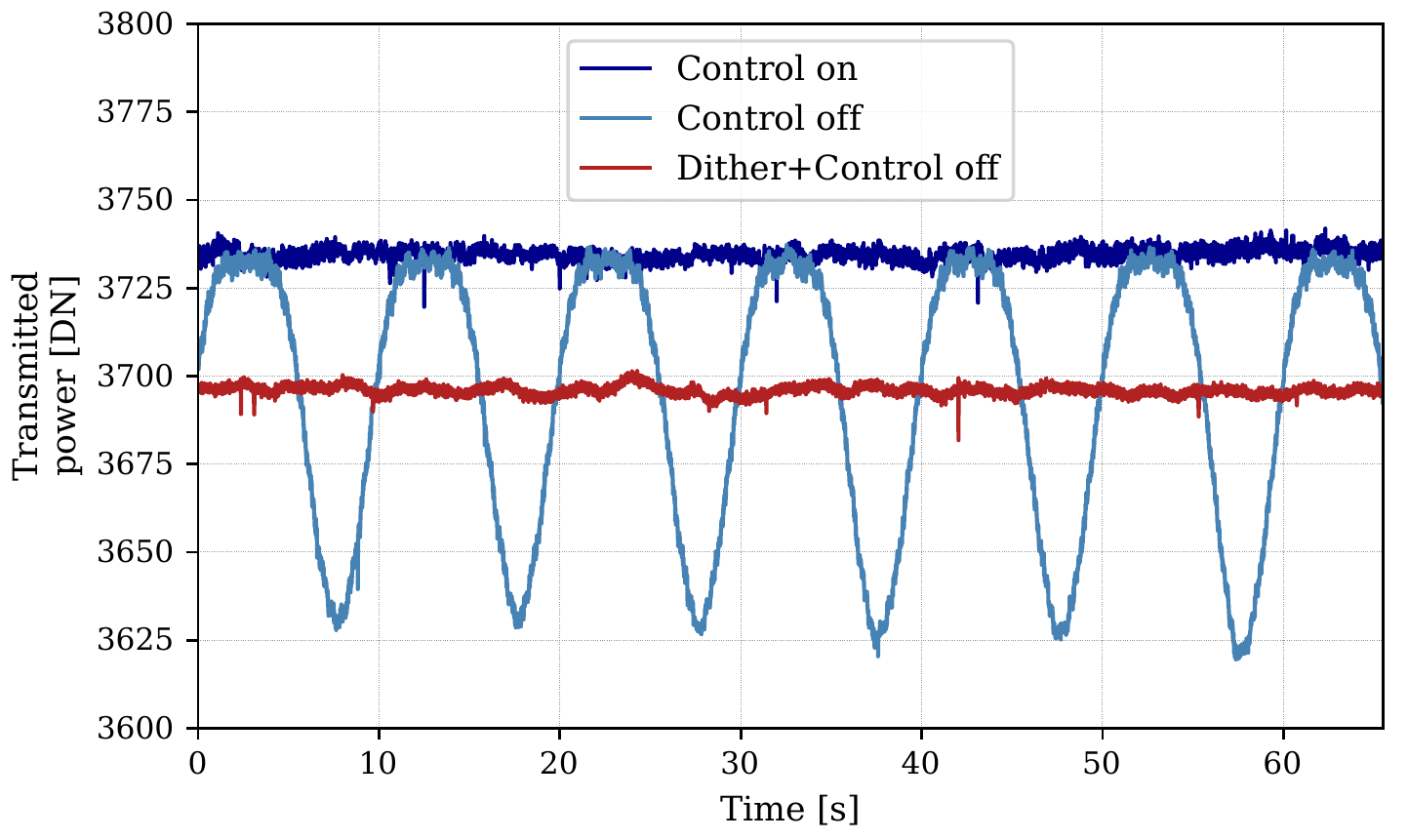} 
\caption{Here we compare three separate 1-minute samples of the optical power on transmission of the cavity. With the alignment and mode-matching control disengaged a 0.1Hz signal was injected into the current of the liquid lenses varying the alignment and mode-matching. The loops were then engaged which successfully compensated for this injected signal.}
\label{fig:dither_locks}
\end{figure}

To test the ability of the system to an evolving wavefront distortions a 0.1Hz dither signal is injected into the liquid lens driver output. Figure~\ref{fig:dither_locks} shows a sample of the time series during this time. Starting with all the loops off and no signal applied, the transmitted power is not optimal. Starting the dither we see the transmitted power peaks. Switching on all the controls loops then suppress the focal length changes keeping the transmitted power optimised.

The fundamental noise limit of our OLC should be photo-electron shot noise on each pixel of the frames used to generate $\mathbf{I}$ or $\mathbf{Q}$~\cite{Cao:20p}. For many sCMOS cameras, the technical noises such as readout and dark noise should be smaller than the shot noise contributions. The single-sided power spectral density of the shot noise in digital numbers is simply given by $2 K \sqrt{N_e}$, where $N_e$ is the total number of photo-electrons contributing to the images $V_0$ and $V_\pi$. This is irrespective of how the images are masked.

\begin{figure}
  \centering
  \includegraphics[width=\linewidth]{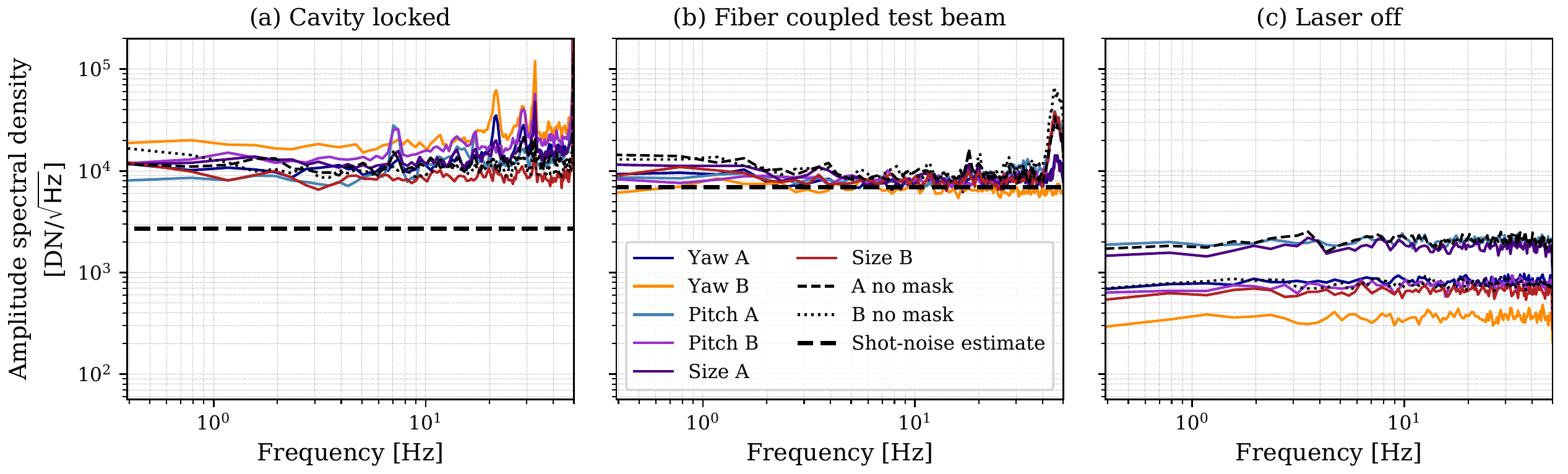}
  \caption{Measured noise spectrums in digital number (DN) for the different degrees of freedom measured by both cameras, 10000 samples were taken at 100Hz in all cases. (a) The cavity length was locked but the alignment and mode-matching loops are off. (b) a test beam coupled from the original laser source via some single-mode fibre, bypassing the actuators and cavity. (c) Full system running but with no laser light on the cameras, this is due to readout and dark noise, as well as any fixed pattern noise which appears differently for each mask type. Also shown for (a,b) is what we estimate the shot noise from the measured photo-electrons to be. Excess noise relative to shot noise is present with the cavity. This is expected to be noise at higher frequencies being aliased down, mostly generated by the liquid lens.}
  \label{fig:asd}
\end{figure}

The noise spectrums for each degree of freedom can be seen in Fig.~\ref{fig:asd}, these were taken over a 100s period.
Fig.~\ref{fig:asd}.a shows the noise present when the cavity is locked. To remove the large DC component of the spectrum the control loops were first engaged to reach the correct operating point, then with the integrators fixed, they were switched off. A flat noise spectrum is seen, however this was not limited by our estimate of shot noise. This excess noise along with many of the visible peaks seen between 10--40Hz that are believed to be a result of noise introduced by the liquid lens. Although the device allows for fast focal length control, it is not a low-noise actuator, significant broadband pitch and yaw noise is introduced between 200-500 Hz, and several resonant features near 1~kHz. The excess noise is suspected to be this noise aliasing down. To test this, we fibre-coupled the laser and bypassed both the actuators and cavity and directly injected the beam into the OLC, the noise spectrum of which are seen in Fig.~\ref{fig:asd}.b. The shot noise estimate for this agrees much better, with some deviation expected due to laser intensity noise, which we do not actively control. Fig.~\ref{fig:asd}.c shows the noise levels in the same state just with the laser light blocked going into the OLC, thus showing the readout and dark noise contributions.

\section{Conclusion}

In this work, we have introduced a new method for imaging RF optical beats at high resolution for performing differential wavefront sensing compared to segmented photodiodes and optical intensity based measurements~\cite{TAKENO20113197,Mah:19}. Using commercially available optical and electrical components, we have shown experimentally that the OLC technique can extract information on both alignment and mode-matching of a beam incident on an optical cavity to optimise alignment and mode-matching. We achieved an alignment and mode-matching to $\approx 99.9$\% by comparing cavity mode scans in the controlled and uncontrolled states.

There are two limiting features for the scheme outlined here. For optimal performance, the camera used must have a fast low-noise readout, with a high bit-depth for subtracting the frames. The camera also requires large well-depth pixels, for maximising the shot noise signal-to-noise ratio. Lastly, we are not able to apply anti-aliasing filters as one would when using a photodiode based readout. This can lead to high frequency noise aliasing and limiting performance. Care must be taken to minimise noise above the sampling frequency of the optical lock-in camera which is discussed in the supplementary materials.

We anticipate such a scheme to find use in next-generation active wavefront control systems for enhancing the performance of gravitational wave detectors and other precision interferometers. This could be for tuning of actuators to correct for low-spatial frequency bulk absorption in optics or point absorbers in optical coatings that have been found to induce higher-spatial frequency distortions~\cite{2101.05828}. A range of low-noise thermal~\cite{ligo_tcs, Kasprzack:13, Wittel:18, beniwal:19, Cao:20c} and mechanical~\cite{Cao:20} actuators have been proposed to combat these issues. Some of these provide actuation at a higher resolution than previously possible, therefore an appropriate sensing scheme, such as ours, would be required for determining how these actuators are affecting the interferometer and provide errors signals for closed loop control systems. 

This project was supported by the Australian Research
Council grant CE170100004.

\section*{Disclosures}

The authors declare no conflicts of interest.

\bibliography{sample}
\end{document}